# Macrostate Parameter - an Econophysics Approach for the Risk Analysis of the Stock - Exchange Market Transactions

**Anca GHEORGHIU**[*]

*Hyperion University of Bucharest, 169 Calea Călăraşilor, 030615, Bucharest, Romania*

**Ion SPÂNULESCU**

*Hyperion University of Bucharest, 169 Calea Călăraşilor, 030615, Bucharest, Romania*

In this paper we attempt to introduce an econophysics approach to evaluate some aspects of the risks in financial markets. For this purpose, the thermodynamical methods and statistical physics results about entropy and equilibrium states in the physical systems are used. Some considerations on economic value and financial information are made. Finally, on this basis, a new index for the financial risk estimation of the stock-exchange market transactions, named macrostate parameter, was introduced and discussed.

## 1. Introduction

Although it has appeared recently, at the beginning of the 21$^{st}$ century, econophysics imposed itself – by the numerous articles, books and communications at international conferences – as an interdisciplinary science, representative and extremely useful to analyse and to model socio-economic systems.

The majority of the studies and works published or communicated at conferences or symposiums present the results of the modeling of the capital markets using mostly mathematical statistics methods and statistical physics methods (see for instance, Levy et al. (2000), Eguiluz and Zimmernann (2000), Drăgulescu (2000 and 2004), Antoniou et al. (2004), Lillo and Mantegna (2000), Mantegna and Stanley (2000), Mantegna et al. (1999), Foley (1994)).

In the last years, many researchers included in their papers some models based on analogies between the economical phenomena and phenomena from other fields of physics such as thermodynamics, electricity, spectroscopy, phase transitions physics, reliability theory etc. ( Johansen et al. (2000), Sornette et al. (1996), Wang Yougui et al. (2004), Gheorghiu and Spânulescu (2004), Gheorghiu (2007)).

---

[*] anca.gheorghiu@gmail.com

In some papers the sharp economic development and even stock-exchange market crashes are assimilated to phase transitions and related to the "critical points" from thermodynamics (see for instance Johansen et al. (2000), Wand Yougui et al. (2004)).

In order to analyse some aspects of the risks in the financial market transactions, we used in this paper the phenomenological thermodynamics methods and the statistical physics results concerning statistical interpretation of entropy and of the equilibrium conditions of physical systems. In the like manner, in the next section some considerations upon the economic value and the information fascicle in the financial markets analysis are made.

Section 3 recalls some considerations upon the volatility and, by correlation with the Section 2 results, a concept of the normalized volatility is proposed and analysed. In the Section 4, using this concept and correlation with Section 2 considerations, on the basis of the analogy between the thermodinamic entropy and the degree of disorder from financial markets, an important index, named macrostate parameter, is proposed. This parameter can be very useful to analyse risks in the stock-exchange market transaction and to predict the expectations of investors.

Finally, in Section 5 the principal conclusions are summarized.

## 2. Economic value and economic information

As it was shown in the previous papers (Gheorghiu and Spânulescu (2004), Gheorghiu (2007)), besides its intrinsic value, characterized by use value and trade value, an informational type value which determines the denomination, role and importance of the product or service, can be ascribed to any good, service or product.

The utility value as well as the exchange value are more tied to the mass, order concept or the construction of things, whereas the own information of a material good having an immaterial aspect like a wave, is not physicaly palpable, therefore it presents a waved (pulsatory) aspect: we either have or not the respective information about the considered product or service. In this vision, if the number of informations is large and concentrated upon some objects or economic "targets" etc., we can speak about an informational fascicle (beam) with dual aspect, similar to a photon beam (from physics) or other elementary particles (electrons etc.) characterized by a determined motion mass, impulse etc.

Taking into account the previous considerations, in the case of systems with a big number of constituents as are the financial markets, money, share number, the price or the volume of the shares transacted on the markets etc., the laws of statistical physics or probability theory and statistical calculus can be applied. So several authors esteem that the shares as well as money or economico-financial informations about the type or prices of the shares are compared to the particles from an informational beeam rather by similitude and not by identification with objects from the physical reality. As it will be shown later, such a representation, used by other researchers too (see e.g. Khrennikov$_1$ (2005), Gheorghiu (2007), Knrennikov$_2$ (2005)), is useful to us to introduce new parameters or indexes to characterize the risk of the transacted shares on the financial markets.

## 3. Simple volatility and normalized volatility

For the financial markets analysis, the proper instruments of technical analysis that can provide valuable information about the evolution of the various transacted assets, are used. Among the instruments of technical analysis, those used the most by investors are simple graphs to indicate the prices evolution or the volumes transacted, the simple volatility or logarithmically expressed volatility, simple averages or Bollinger bands and lesser other indexes, stochastic oscillators etc., which are approached by the specialists or financial analysts.

For volatility, the relation that gives the shares price difference at two successive moments is usually used:

$$Vol = p_t - p_{t-1} \tag{1}$$

where $t$ represents the present time, and $t-1$ is the time at the previous moment, separed from $t$ by the time unity (minute, hour, day, month, year etc.), as well as the logarithmic expression:

$$Vol = \ln(p_t) - \ln(p_{t-1}) \tag{2}$$

which allows the graphic representation for longer time periods.

From the stock-markets analysis it is sometimes established that, altough some assets with high prices are well appeciated, having an increasing tendency of the price, they are characterized by a diminished liquidity because of smaller transaction volumes, leading to ampler price oscillations, i.e. more risky for investors. Contrarily, the assets with a more reduced price that can attract the investors, consequently being able to determine large transacted volumes, can inspire some confidence on the market although the shares do not have a corresponding good evolution (having many price oscillations and price corrections etc.)

In the first case mentioned it can be said that such a share of high value, has a much more inertia on an increasing/decreasing tendency of the transacted price and at the same time, by reduced liquidity, is much more risky, as will compare a lump pile with a sand hillock. Even if the volume should be the same, the effects are very different. From this point of view the product price × transacted volume can be assimilated with to impulse of a particle defined by the product of mass $m$ and speed $v$:

$$p = m \cdot v \tag{3}$$

from the elementary physics.

For a more complete understanding of the share evolution from the point of view of the price and transacted volumes, the product **price × transacted volume** can be assimilated to the impulse of a particle (which symbolize the respective financial information) defined by the product $pV$ similar to the impulse of a particle-information defined by a relation of (3) type.

Such an index can delivers ampler useful information regarding the "inertia" degree or stability of an asset (shares, financial instruments etc.) than the price, $p$, or the transacted volume, $V$, taken separately.

In the like manner, it can be also considered other parameters that result from combinations of the two entities of the type: price/volume or volume/price etc., all these combinations can be simbolically marked by parameter $a_t$.

Considering the product price × volume, $a = pV$, we can define the normalized volatility as:

$$Vol_n = \frac{p_t V_t - p_{t-1} V_{t-1}}{p_{t-1} V_{t-1}}, \qquad (4)$$

where:

$p_t$ is the closing price from the day $t$;

$V_t$ is the number (volume) of transacted shares in the day $t$;

$V_{t-1}$ is the number (volume) of transaction shares in the day $t$-1;

$P_{t-1}$ is the closing price from the previous day $t$-1.

The normalized volatility represents a powerful index of the share's state (condition or status etc.) compared with the previous day and meanwhile an information refering to the investor perceiving with respect to the asset and theirs expectations toward the investment in the respective share (company) from a moment to another. This index will be used in the next section to introduce and analyse the macrostate parameter - a new index very useful for the risk estimation in the financial market transactions.

## 4. The macrostate parameter of the financial market transactions

As we mentioned above, all the informations from the financial field can be assimilated to the particles of a gas of impulses $p = mv$ (see relation 3) confined in a precinct ("financial boiler") that is the very capital market (spot markets, forward markets etc.). In this situation it is plausible enough to apply the same principles, laws and results from thermodynamics, kinetic-molecular or statistical physics to describe the assembly of particle states – called microstates – in which the particles that symbolise the information about the shares (or other financial instruments) from the virtual precinct can exist at various moments. Every "particle-information" contained in the financial boiler (virtual precinct) is characterized – in a first phase – by the product price × volume of transacted shares, i.e. by the parameter $a = pV$ as we have seen in the previous section.

After a determined time, as a result of the succession of a numerous microstates which appear because of the agitation and the mixture of the constituent particles, the system reach an equilibrium state which is a **macrostate** that can be described by measurable macrostate parameters (Toda et al. (1992)).

By financial (or economic) macrostate we understand the assembly of information and decisions materialized in the share prices and the transacted volume (individual and for all the day, hour or minute of transaction) when we refer to an emitter quated, or to the quote of the stock-market index, if the market is analysed in totality or for one section of it.

So, if we consider $a = pV$ parameter, a microstate for the capital market is given by the assembly of the price and share's volume situation at a very moment $t$. If this microstates is altered by the transaction of a single investor, for example, who buys or sells a single share, the assembly of microstates will be modified, resulting a new picture of the stock-market situation, i.e. a new microstate. These changes are practically infinitely numerous (it can be a big number of transactions) so the microstates number should be extraordinary big and can be interpreted and statistically evaluated by a similar formula like the one given by L. Boltzmann for the microstates of a thermodynamic system (mix of gases) which defined the entropy of a thermodynamic system (Gheorghiu (2007), Toda (1992)):

$$S = k \ln W. \qquad (5)$$

In expression (5), $k = 1.380.662 \times 10^{-23}$ J/K represents the Boltzmann's constant, and:

$$W = \frac{n!}{n_1! n_2! n_3! \ldots} \qquad (6)$$

is the thermodynamics probability to realize a microstate of the system.

Starting from the definition (5) of entropy $S = k \ln W$ from thermodynamics, we can introduce a similar parameter named macrostate parameter, for the financial markets:

$$P_M = k_B \ln W_B \qquad (7)$$

where $W_B$ represents a probability in succeeding a new microstate of the stock-market and $k_B$ is a constant which is specific for that stock-exchange market and for that type of transacted share.

On the other hand, the entropy $S$ is tied to the thermodynamic temperature, $T$, by a relation like (Gheorghiu (2007), Toda (1992)):

$$dS = \frac{\delta Q}{T} \qquad (8)$$

or for the finite variations, by the relation:

$$S = \frac{\Delta Q}{T} \qquad (9)$$

that is to say that the thermodynamical entropy is proportional to the reverse of the temperature $T$.

Similarly, for the financial markets, to abstract a proportionality factor $\Delta Q$, the reverse of $P_M$ represents the stock-exchange market temperature $T_B$, which is also an important macrostate parameter for the financial markets analysis (see relation 9):

$$T_B \cong \frac{1}{P_M}. \qquad (10)$$

If we imagine a virtual precinct, characterized by the parameters of a momentary economic state (time, index value at the end of the transactional day and the volatility defined as the index value variation from a day to another one), we can "visualize" the state of a given period, as it can be seen in figure 1 of the romanian stock-market indexes (for data series from the creation of the respective market index up to December 2007).

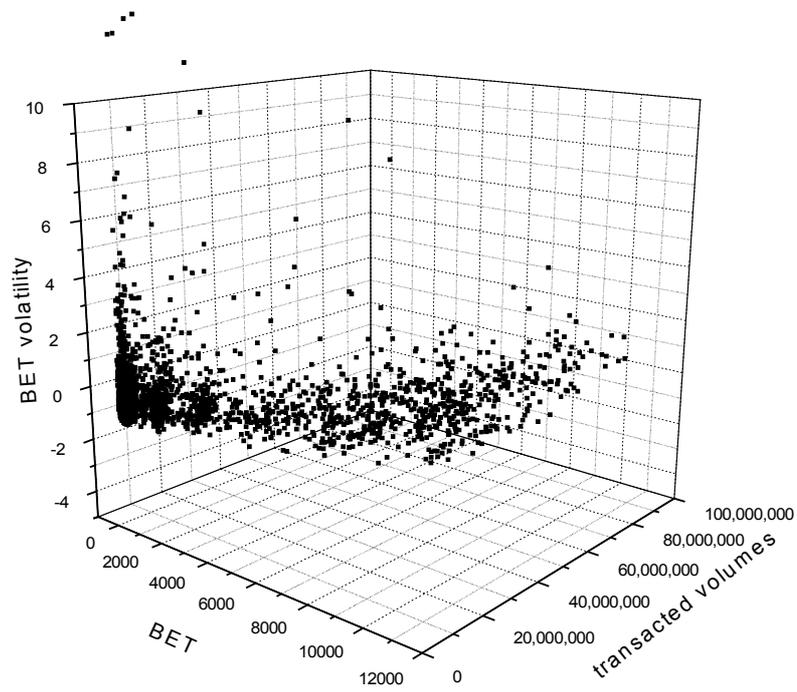

a) BET

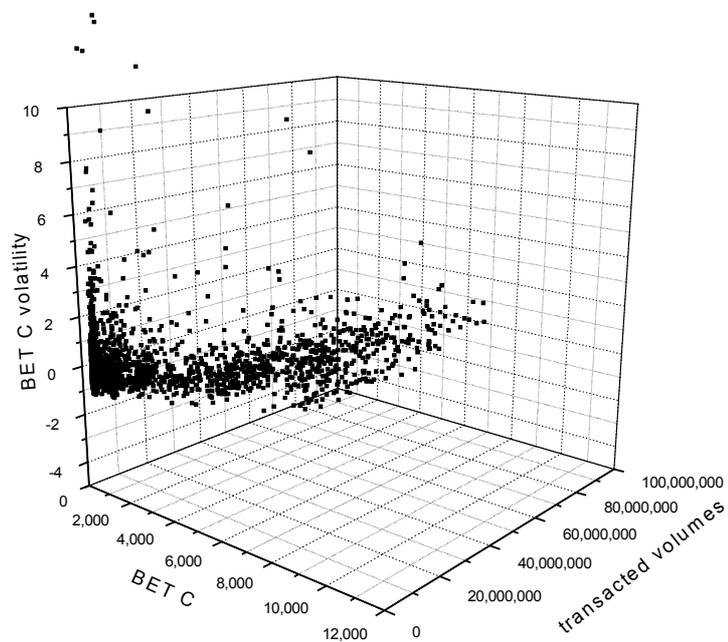

b) BET-C

**Figure 1.** "Thermal" precincts of the romanian stok-market indexes (Gheorghiu (2007)).

Considering the displacement in the linear direction of some imaginary particles, the microstate assembly of all the particle mouvement reflects the "thermal stir" state of particles contained in the precinct. The bigger the displacement speed $v$ is, the more, the impulse $p = mv$ of these particles will increase, as well the precinct temperature and the system energy will increase. As it was mentioned before, for the informational precinct (i.e. "financial boiler"), the product $a = pV$, is similar to the impulse $p = mv$ of the material microparticles which, in the case of the financial virtual precinct, is closely tied to the volatility, and to the normalized volatility respectively, defined by relation (4).

So, in our considerations, instead of simple "impulses" of gas particles whose assembly's motion determines the temperature of the medium of the precinct, we will use the normalized volatility (see rel. (4)):

$$Vol_n = \frac{a_t - a_{t-1}}{a_{t-1}}. \tag{11}$$

A microstate of the stock-exchange market is given by the assembly of total prices and trade volumes for transactioned shares at a given moment $t$. The addition of the share's information microstates can be represented like an assembly of the impulses of the particles in a gas which gives the dimensions of the macrostate, that means, the temperature of the environment or of the precinct which contains the studied "gas". The "virtual precinct" that we have proposed is a three – dimensional one, with the following coordinates: price, transacted volume and time period $t$. In other words, for the temporal dimension taken into account, we can determine the microstates of each day of transaction and a macrostates parameter that can be defined by the sum of all microstates divided by the totality of their number, $N$, that could be similar to the financial entropy $P_M$.

Consequently, the sum of all microstates, defined by the normalized volatility divided by number of microstates, $N$, gives the dimension of the agitation (disorder) on the market for a given period, as a macrostate parameter:

$$P_M = \frac{\sum \text{normalized volatility}}{N} = \frac{1}{N} \sum_{t=1}^{N} \frac{a_t - a_{t-1}}{a_{t-1}} \tag{12}$$

where $a_t$ symbolise the price $\times$ volume product.

Comparing the formula (12) with the relation (7) for entropy, we can also write the equality:

$$P_M = \frac{1}{N} \sum_{t=1}^{N} \frac{a_t - a_{t-1}}{a_{t-1}} = k_B \ln W_B = S_e \tag{13}$$

where $S_e$ represents the economic entropy being determined and equal to the macrostate parameter $P_M$.

If in the relation (13) we can consider the factor $k_B$ as being given by $1/N$, it results that the expression $\ln W_B$ can be identified (and evaluated) by the sum of the characteristic normalized volatilities for the system microstates and defined by the relation (11).

As it was mentioned, a thermodynamic system is wholly described by dint of two macroscopic parameters. A macroscopic parameter represents the statistical value of all microscopic states at a given moment $t$ or under certain conditions (of volume, impulse etc.). Such a parameter is the macrostate parameter $P_M$ defined by the relation (12), as well as its reverse, the financial market temperature, $T_B$ (see also rel. 10):

$$T_B = \frac{1}{P_M} = N \sum_{t=1}^{N} \frac{a_{t-1}}{a_t - a_{t-1}} \qquad (14)$$

where $a_t = p_t V_t$.

To illustrate this, in figure 2 the "economic virtual precincts" of some romanian petroleum companies performances, currently quoted at Romanian Bourse for 4.07.2004-23.11.2006 period are represented. The normalized volatility is represented as price as well as volume function i.e. as $a_t = f(p, V)$ function.

On the basis of volatility data, price and volume, in figure 3,a the values for $P_M$ and $T_B = \frac{1}{P_M}$ given by (13) and (14) formulas, where $a_t = p_t V_t$ are represented. In the figure 3,b the same results (under histogram form for $P_M$ parameter), much more suggestive for interpretation and analysis, are represented.

To be able to appreciate the perception degree of the investors, besides the four petroleum companies analysed in the figure 2, we applied the same calculus for two shares situated at the extremities of the stock-exchange market i.e. SOF, which represents the symbol of Sofert Bacău, a society with weak results for which it was declared the payment incapacity and SIF5, symbol of SIF Oltenia, from the financial field, which in 2004-2006 period, appreciated itself with more 200%, very transactioned and interesting for the speculator's portfolio, but as well for the investors for the dividends annualy granted. The comparative results are ilustrated in the figure 3,a and 3,b respectively.

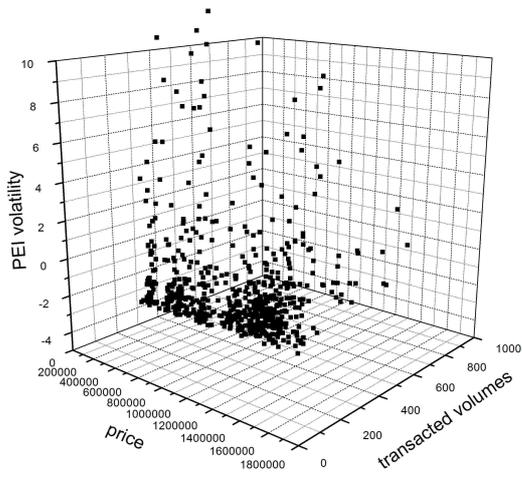

a) PEI price × transacted volumes × volatility

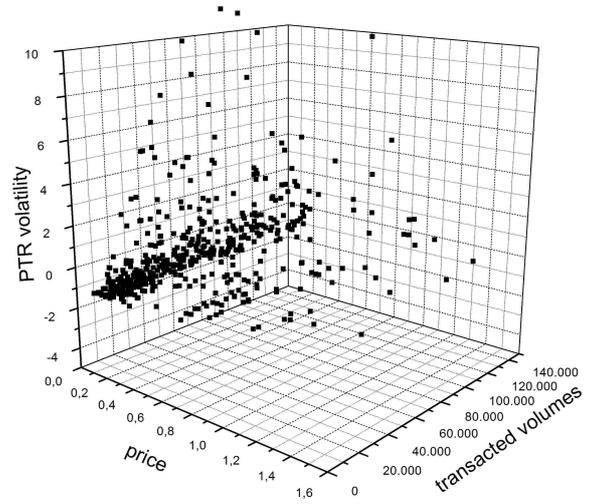

b) PTR price × transacted volumes × volatility

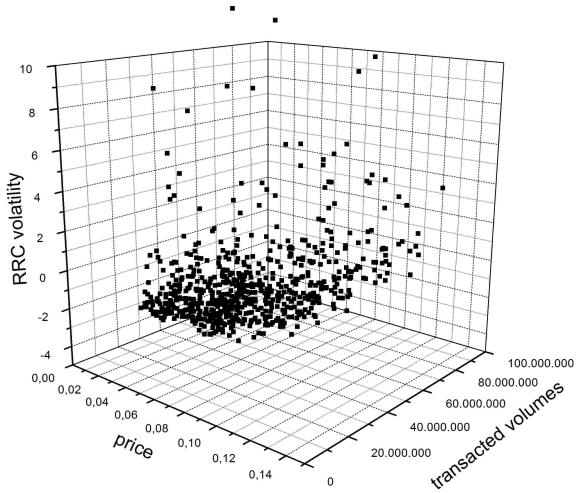

c) RRC price × transacted volumes × volatility

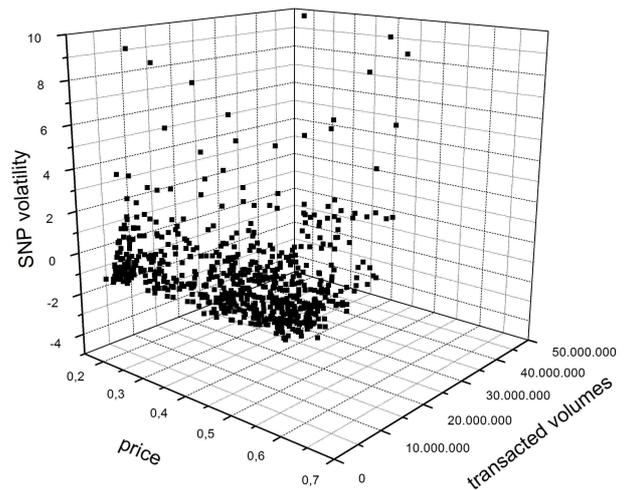

d) SNP price × transacted volumes × volatility

**Figure 2.** Economic virtual precincts of the performances of some romanian oil companies quoted at BVB (04.07.2004 – 23.12.2006) (Gheorghiu (2007))**.**

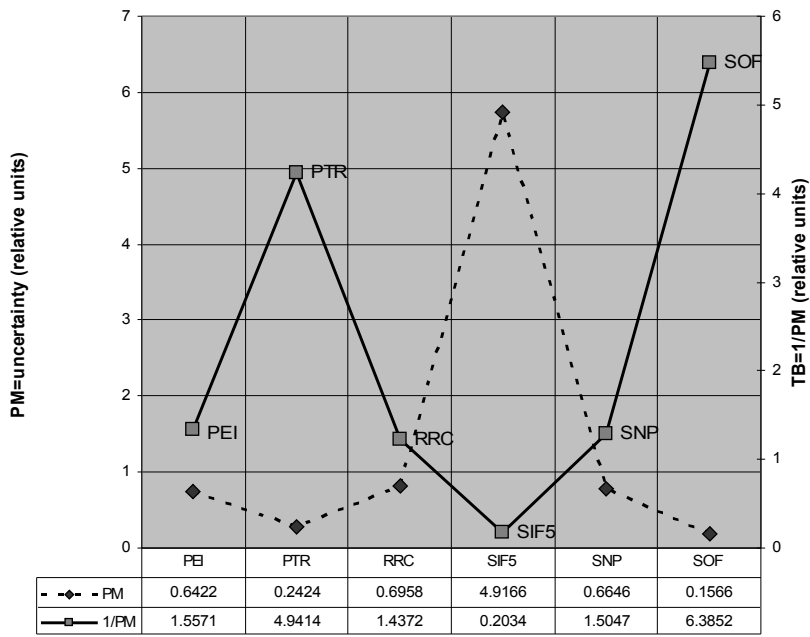

a)

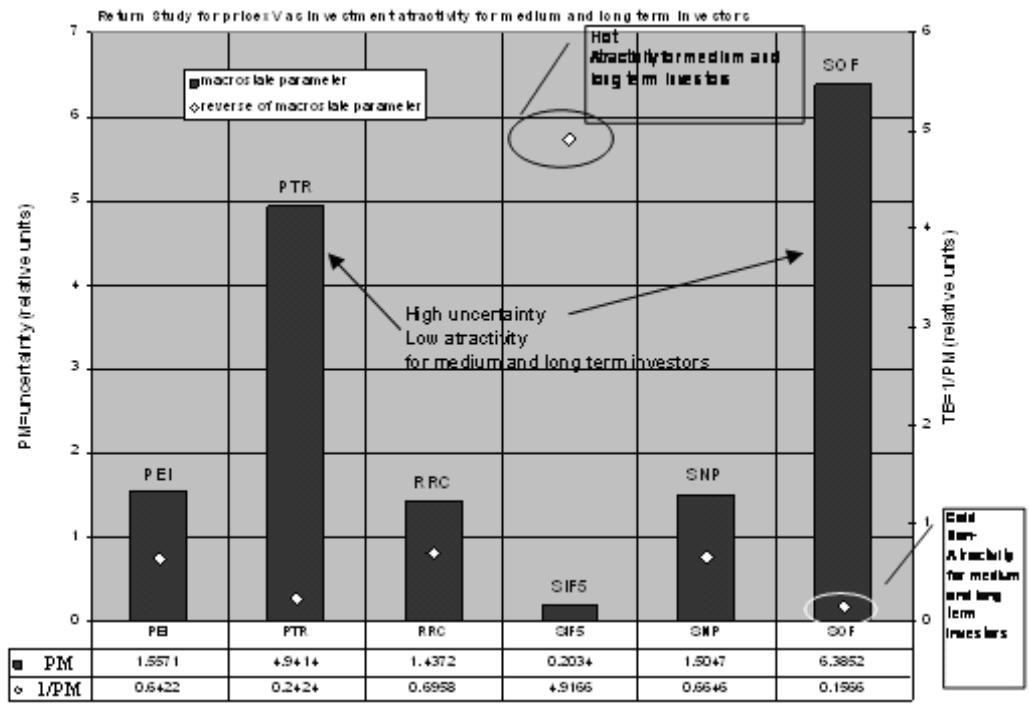

b)

**Figure 3.** a) The macrostate parameters $P_M$ and $1/P_M$ for the some companies transacted on Romanian Stock-exchange market (BVB); b) The same results for the quoted companies, the macrostate parameter $P_M$ being represented by histogram forms.

As it was specified above, in all the analysed cases, a share is characterized by the doublet price × transacted volume, state parameter $P_M$ and $T_B = 1/P_M$, being calculated taking into account the product $pV = a$ (see (13) and (14) relations).

As it was shown by the relations (10) and (13), the macrostate parameter $P_M$ defines the economic entropy $S_e$, and the reverse of the macrostate parameter, $1/P_M$, represents the economic (financial) „temperature" $T_B$ of the capital market, for the chosen criteria (parameter $a = pV$) and the analysed companies (Gheorghiu (2007)).

Let us argue the aspects illustrated in figures 3,a and 3,b. Under the investors' certitude, the investors are strongly attracted to SIF5 (the investment fidelity on medium and long periods), what we knew, but what we did not know is just the high "temperature" this share proposed to the market (see the circle round the point up on figure 3,b). The shares PEI, SNP and RRC are situated on moderate market equilibrium positions, unlike of PTR, which although has a financial good situation is affected of low liquidity and a reduced visibility in the markets, which generate an anormal "thermal" agitation. And sulrey, at antipodes, SOF, the "coolest" share and the same time transacted on large volumes because of a very low price, of a lack of investment culture of some players on the market and of the presence in zone of imprudent speculators.

If we look at charts where the diagrams of the macrostate parameter $P_M$ is represented (figures 3,a or 3,b), we can directly see how it can be used. Its value is showing us the degree of uncertainty for uncertain shares although they have a large volume of trading (see SOF or even PTR cases). We can conclude that with this low interest shares, the entropy is bigger mostly due to the chaos induced by the market because of low prices and large transacted volumes despite of strong uncertainty and lowly liquidity.

To be sure that the precedent reasonings are correct and that we can introduce a new technical index for the portfolio investor on medium and long periods, we extended the study on the same emittent quoted on four markets namely on a spot romanian market (BVB), on forward market (BMFSMS) at two different terms of payment and on a foreign market (Wiener Boerse). The sole emittents, in the category selected by us (petroleum sector) which fulfilled this condition at the moment of analysis were RRC and SNP, respective DE_RRC and DE_SNP and OMV respectively.

The selected time for this analysis was one month, namely January, 2006. The analysis results are given in figure 4.

By using the macrostate $P_M$ and respective $T_B = 1/P_M$ parameters, it can be immediately seen which are the attractive shares (Fig. 4) and how intervenes the uncertainty factor, which in this situation (future market) acts in more varied ways.

The fact that $T_B = 1/T_M$ takes low value in DE_RRC_JUN can be explained by the fact that the transaction of the contracts with term of payment on 30 June just begun (the sampler period being January 2006), the traders are prudent and small quantities of shares are transacted in order to identify the probable trend of the market, therefore they are very volatile and risky, being transacted on low volumes (some time are not transacted). On the contrary, DE_RRC_MAR and DE_SNP_MAR have been transacting for three months (from 1-October 2005) and the market established a trend.

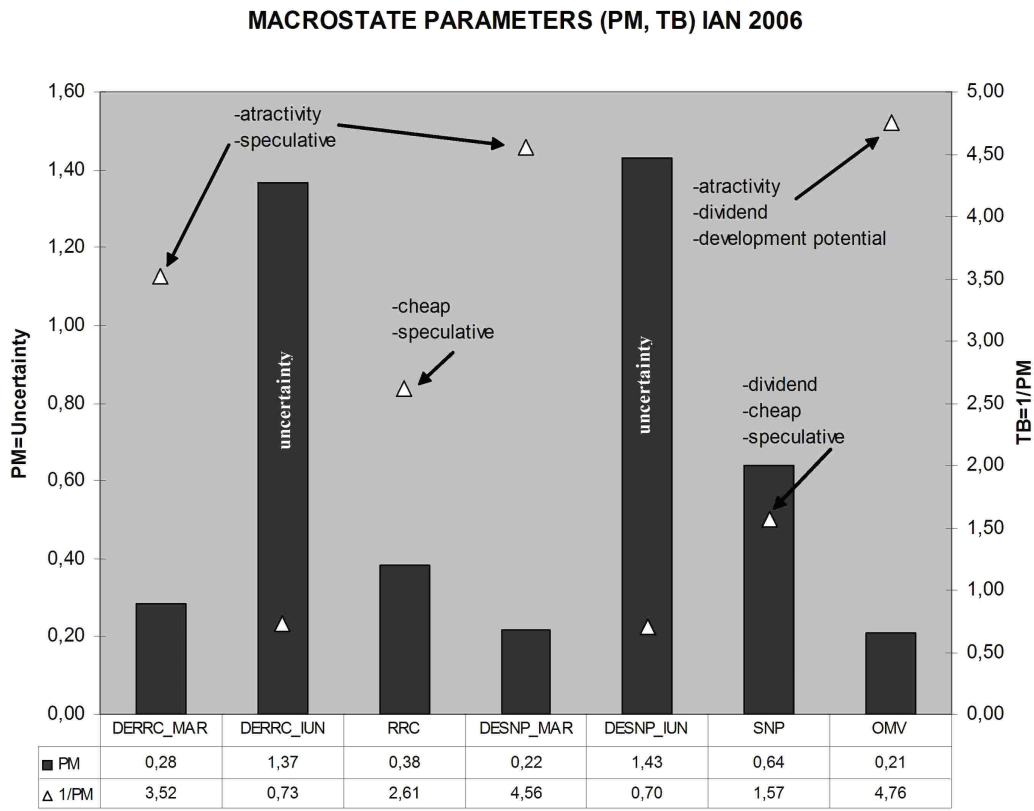

**Figure 4.** The macrostate parameters for RRC and SNP on the spot (BVB) and the futures (BMVFMS) markets and OMV on the spot Austrian (Wiener Boerse) market.

For OMV (Austrian company), from figure 4, we can see that macrostate parameter $P_M$ is much lower compared to RRC and SNP, which denotes a much diminished entropy, i.e. a smaller uncertainty degree for this emitent, in other words the investors "know" better the company potential and its development plans.

From what was shown, it is seen that, practically, the macrostate parameter is a measure of the uncertainty from an economic point of view and a measure of the disorder (entropy) from an econophysics point of view. In other words the economic uncertainty can be measured by the macrostate parameter, which is the entropy equivalent from physics.

Taking in account the considerations made in this section we extended our study on the 40 companies quoted in the I and II categories of the BVB (Romanian stock - exchange market) and applying the methodology described above, using the macrostate parameter, two investment risk scales for the companies enlisted during 2006 and 2007 years were established (Fig. 5 and Fig.6). The financial results of the companies corroborate with their hierachy on the macrostate parameter value for the 2006 and 2007 years verified the theoretical considerations of foundation of the econophysics model for the risk estimation in the stock-exchange markets.

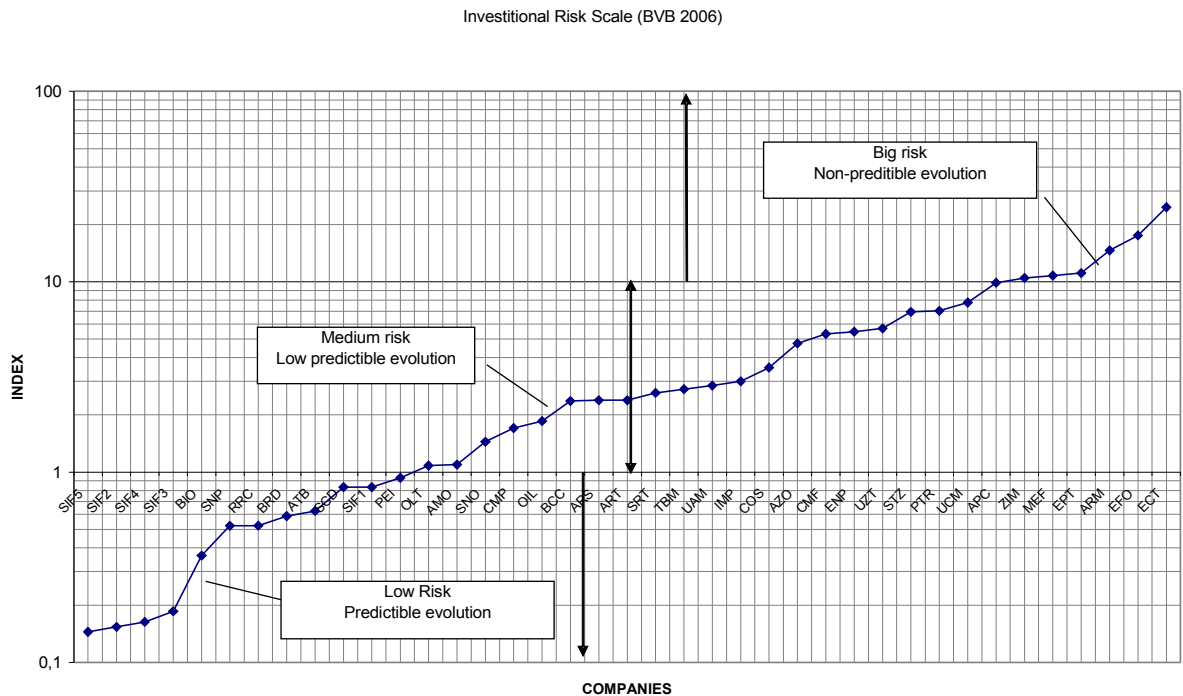

**Figure 5.** Investitional risk scale for the 40 romanian companies quoted at Bucharest stock - exchange market (BVB) during 2006 year.

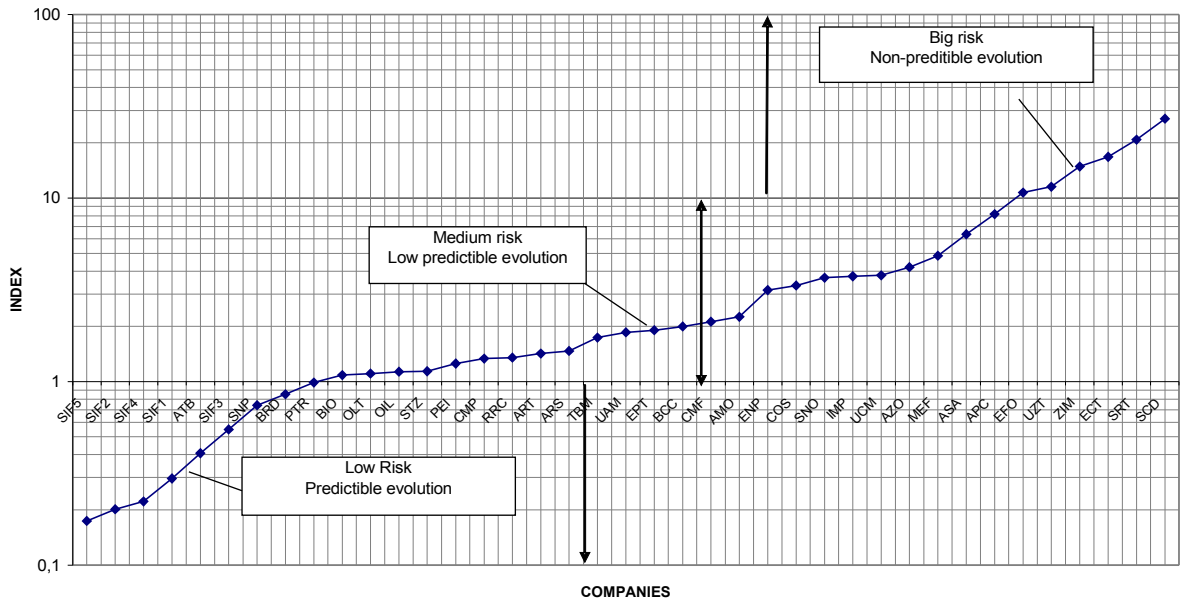

**Figure 6**. Investitional risk scale for the 40 romanian companies quoted at Bucharest stock - exchange market (BVB) during 2007 year.

# 5. Conclusions

Considering the information about stock-exchange market like a financial boiler, on the basis of the analogy of the thermodynamic entropy and the disorder degree from the capital (financial) markets, a new index for the risk estimation and evolution of shares quoted on regularly markets, named macrostate parameter, was introduced.

From the econophysics point of view, the macrostate parameter $P_M$ is a measure of the disorder (entropy) from the capital market, being similar to the thermodynamic entropy from physics. Its size shows directly the uncertainty degree from the economic (financial) point of view for various transacted companies, being able to constitute a measure of the attractive degree on the medium or long term. This method can be applied on multiple financial exercises or on a single fiscal year. As a conclusion, it results that $P_M$ is a strong parameter for the appreciation of the investors if to take or not into consideration for their portofolio one or other companies.

The macrostate parameter "smoothes out" the useful information, dismisses the "parasite noise" and put into evidence the value of risk and the real appetite of the investors towards the analysed companies.

The macrostate parameters are intensive and deliver synthetical informations to the investor on the regulated markets, on the basis of which he can build scales of risk and investitional uncertainty for the transacted companies on the financial markets.